\documentclass[12pt,aps]{revtex4}

\usepackage{graphicx}
\usepackage{bbm}
\usepackage{amssymb}
\usepackage{amsmath}

\newcommand{\nc}{\newcommand}
\nc{\be}{\begin{equation}}
\nc{\ee}{\end{equation}}
\nc{\bea}{\begin{eqnarray}}
\nc{\eea}{\end{eqnarray}}
\nc{\nn}{\nonumber}
     
\begin{document}

\title{ Gravitational Wave Production by Collisions: More Bubbles}

\author{Stephan J. Huber$^{(1)}$ and Thomas Konstandin$^{(2)}$}

\email[]{s.huber@sussex.ac.uk}
\email[]{konstand@ifae.es}

\affiliation{$(1)\,$ Department of Physics and Astronomy, Sussex, Falmer, Brighton, BN1 9QH, UK}

\affiliation{$(2)\,$ IFAE, Universitat Aut\`onoma de Barcelona, E-08193 Bellaterra, Barcelona, Spain}

\date{\today}

\begin{abstract}
We reexamine the production of gravitational waves by bubble
collisions during a first-order phase transition. The spectrum of the
gravitational radiation is determined by numerical simulations using
the "envelope approximation".  We find that the spectrum rises as
$f^{3.0}$ for small frequencies and decreases as $f^{-1.0}$ for high
frequencies. Thus, the fall-off at high frequencies is significantly
slower than previously stated in the literature. This result has
direct impact on detection prospects for gravity waves originating
from a strong first-order electroweak phase transition at space-based
interferometers, such as LISA or BBO. In addition, we observe a
slight dependence of the peak frequency on the bubble wall velocity.
\end{abstract}

\maketitle

%
%

\section{Introduction}

Colliding bubbles in a first-order phase transition constitute one
possible source of stochastic gravitational wave (GW)
radiation~\cite{Witten:1984rs, Hogan}. If the electroweak phase
transition is strongly first-order, for instance, the kinetic energy
stored in the Higgs field and the bulk motion of the plasma is
partially released into gravity waves. This happens mostly at the end
of the phase transition, when collisions break the spherical symmetry
of the individual Higgs field bubbles.  This possibility was
systematically analyzed in a series of papers~\cite{Kosowsky:1991ua,
Kosowsky:1992rz, Kosowsky:1992vn, Kamionkowski:1993fg}. The first
simulation~\cite{Kosowsky:1991ua} consisted hereby of the full scalar
field dynamics of two bubbles in vacuum, where the essential
observation was made that the emitted radiation depends only on the
gross features of the problem, namely the kinetic energy stored in the
uncollided bubble regions. This observation is the basis of the
so-called envelope approximation that opened up the possibility of
simulating phase transitions with a large number of bubbles. This was
subsequently exploited in refs.~\cite{Kosowsky:1992rz,
Kosowsky:1992vn} and further refined for a thermal environment in
ref.~\cite{Kamionkowski:1993fg}.

In the case of only two colliding bubbles, the spectrum decreases as
$f^{-1.8}$ for high frequencies. But this result might be special to
the case of two bubbles, where the collision never finishes, which
makes the introduction of a time cutoff function
mandatory~\cite{Kosowsky:1991ua}. The situation is different if a
realistic phase transition with a large number of bubbles is
simulated. There were hints in refs.~\cite{Kosowsky:1992rz,
Kosowsky:1992vn, Kamionkowski:1993fg} that the spectrum of
multi-bubble simulations might be more flat than in the two bubble
case, but the numerical accuracy prohibited a conclusive statement.
As a result, the frequency fall-off of the two bubble case is still
being used in the present day literature~\cite{Grojean:2006bp, KonHub,
Kahniashvili:2008pf}.

The aim of the present work is to reexamine the generated spectrum of GWs
by simulating a phase transition with a large number of bubbles, making
use of the aforementioned envelope approximation. Compared to
ref.~\cite{Kamionkowski:1993fg}, the numerical accuracy will be
considerably improved, and a larger portion of the spectrum will be determined to
allow for a careful analysis of the high frequency behavior.

\section{Determination of the GW spectrum}

The fundamental quantity that enters the determination of the
gravitational radiation are the spatial components of the
stress-energy tensor $T_{ij} ({\bf x}, t)$.  For a thermal phase
transition it consists of the scalar field part and the plasma
contribution.  The total energy radiated into a direction $\hat {\bf
k}$ is then given by~\cite{Weinbergbook}
\be
\label{eq_weinberg}
\frac{dE_{GW}}{d\omega d\Omega} = 2 G \omega^2 \Lambda_{ij,lm} (\hat {\bf k})
T^*_{ij} (\hat {\bf k}, \omega) T_{lm} (\hat {\bf k}, \omega),
\ee
where $T_{ij} (\hat {\bf k}, \omega)$ denotes the stress-energy tensor
in Fourier space
\be
\label{eq_Tij_Fourier}
T_{ij} (\hat {\bf k}, \omega) = \frac{1}{2\pi} \int dt \, e^{i\omega t}
\int d^3x \, e^{- i \omega \hat {\bf k} \cdot {\bf x}} 
\, T_{ij} ({\bf x}, t),
\ee
and $\Lambda$ is the projection tensor for the transverse-traceless
part
\be
\Lambda_{ij,lm} (\hat {\bf k}) = 
\delta_{il} \delta_{jm} - 2 \hat {\bf k}_j \hat {\bf k}_m \delta_{il}
+ \frac12 \hat {\bf k}_i \hat {\bf k}_j \hat {\bf k}_l \hat {\bf k}_m
- \frac12 \delta_{ij} \delta_{lm} 
+ \frac12 \delta_{ij} \hat {\bf k}_l \hat {\bf k}_m
+ \frac12 \delta_{lm} \hat {\bf k}_i \hat {\bf k}_j.
\ee
Eq.~(\ref{eq_weinberg}) is derived in the wave zone approximation that
is well justified in the present case.

The key idea of the envelope approximation~\cite{Kosowsky:1992vn} is
that the GW production does not depend on the details of the evolution
of the scalar field in the region of the collided bubbles, but rather
on the gross features of the problem, namely the shape of the
uncollided bubble walls.  Denoting the times and positions of the
nucleated bubbles by $t_n$ and ${\bf x}_n$, this turns
eq.~(\ref{eq_Tij_Fourier}) into
\be
\label{Tij_a}
T_{ij} (\hat {\bf k}, \omega) = \frac{1}{2\pi} \int dt \, e^{i\omega t}
\sum_n \int_{S_n} d\Omega \,
\int dr \, r^2   e^{- i \omega \hat {\bf k} \cdot ({\bf x}_n + r \hat {\bf x})} 
\, T_{ij,n}(r, t),
\ee
where $S_n$ denotes the uncollided region of the $n$th bubble and
$T_{ij,n}$ its stress-energy tensor. If the phase transition proceeds
by detonation, the stress energy is concentrated in a shell of bulk
motion that is thin compared to the bubble radius, thus motivating the
thin-wall approximation~\cite{Kamionkowski:1993fg}
\be
4 \pi \int dr \, r^2   
e^{- i \omega \hat {\bf k} \cdot ({\bf x}_n + r \hat {\bf x})} 
\, T_{ij,n} (r, t)
\approx  \frac{4 \pi}{3} 
e^{- i \omega \hat {\bf k} \cdot ({\bf x}_n + R_n \hat {\bf x})} 
\, \, \hat {\bf x}_i \hat {\bf x}_j \,  R_n^3 \, \kappa \, \rho_{vac},
\ee
with $R_n(t)$ the size of the $n$th bubble and $\rho_{vac}$ the
difference in energy density between the true and the false vacuum (we
use this terminology even though in the presence of a thermal bath the
latent heat is the relevant quantity).

The efficiency factor $\kappa$ determines how much of the
vacuum energy is transformed into kinetic energy of the bulk fluid
instead of reheating the plasma inside the bubble. This coefficient
can be determined as a function of the parameter $\alpha$ that is defined
as the ratio between the vacuum energy and the total energy stored in
radiation~\cite{Kamionkowski:1993fg}
\be
\kappa(\alpha) = \frac{1}{1 + 0.715 \alpha} 
\left[ 0.715 \alpha + \frac{4}{27} \sqrt{\frac{3\alpha}{2}}\right],\quad 
\alpha = \frac{\rho_{vac}}{\rho_{rad}}.
\ee
The last ingredient from the phase transition is the velocity of the
bubble wall $v_b$. For very strong phase transitions it is given
by~\cite{Steinhardt:1981ct}
\be
\label{eq_vb}
v_b = \frac{\sqrt{1/3} + \sqrt{\alpha^2 + 2\alpha/3}}{1 + \alpha}.
\ee
In fact, eq.~(\ref{eq_vb}) is not true in general. In phase
transitions, there exists a larger class of detonation solutions, as
discussed in ref.~\cite{Laine:1993ey}. Therefore eq.~(\ref{eq_vb})
gives only a lower bound on the wall velocity and we will treat $v_b$
as a free parameter in our analysis.

Following the above considerations, eq.~(\ref{Tij_a}) can be rewritten as
\bea
T_{ij} (\hat {\bf k}, \omega) &=& \kappa \rho_{vac} \, v_b^3 \, 
C_{ij}(\hat {\bf k}, \omega),  \label{eq_Tij} \\ 
C_{ij}(\hat {\bf k}, \omega) &=& \frac{1}{6 \pi} \sum_n 
\int dt \, e^{ i \omega ( t - \hat {\bf k} \cdot {\bf x}_n )} 
(t-t_n)^3 \, A_{n,ij}(\hat {\bf k}, \omega),  \label{eq_Cij} \\
A_{n,ij}(\hat {\bf k}, \omega) &=&
\int_{S_n} d\Omega \,\, e^{- i \omega v_b (t-t_n) \hat {\bf k} \cdot \hat {\bf x}} 
\hat {\bf x}_i \hat {\bf x}_j \label{eq_Aij}. 
\eea
An efficient way of evaluating these integrals numerically is
presented in the Appendix.

Before we discuss the simulation of the phase transition, we comment
on the so-called quadrupole approximation that is given by the limit
${\bf k\cdot x}\to0$ or
\bea
C_{ij} (\omega) &=& \frac{1}{6 \pi}
\sum_n \int dt \, e^{ i \omega t } 
 (t-t_n)^3 \, A_{n,ij}(0, t).
\label{def_quad}
\eea
This approximation becomes exact in the limit of small frequencies. For
large frequencies, it turns out that this substitution in the case of
two bubbles overestimates the resulting GW production by more than one
order of magnitude; additionally, the peak frequency is shifted to
larger values what was first noticed in the simulation of the scalar
field evolution of two single bubbles in
ref.~\cite{Kosowsky:1991ua}. This result is rather surprising, since
{\it e.g.} in electrodynamics with a localized source, the quadrupole
approximation generally underestimates the full result. However, in
the present case the assumption of localized sources is not applicable
such that the quadrupole approximation can overestimate the full
result as detailed in ref.~\cite{Kosowsky:1991ua}.

To model the phase transition, we assume that the nucleation
probability per volume and time is given by
\be
\label{nuc_prob}
P = P_0\exp(\beta (t-t_0)).
\ee
The parameter $\beta^{-1}$ is approximately the duration of the phase
transition. Since $\beta$ is the sole dimensionful parameter of the
phase transition, the fraction of energy liberated into gravitational
wave radiation per frequency octave is~\cite{Kamionkowski:1993fg}
\be
\Omega_{GW*} = \omega \frac{dE_{GW}}{d\omega } \frac{1}{E_{tot}} = 
\kappa^2 \left( \frac{H}{\beta} \right)^2 
\left( \frac{\alpha}{\alpha + 1} \right)^2 \, \Delta( \omega / \beta, v_b),
\ee
where we used the definition of the Hubble constant
\be
H^2 = \frac{8 \pi G \rho_{tot}}{3} = \frac{8 \pi G (\rho_{vac} + \rho_{rad} )}{3},
\ee
and defined the dimensionless function $\Delta$ as
\be
\Delta( \omega / \beta, v_b ) =  \frac{\omega^3}{\beta^3} 
\frac{3 v_b^6 \beta^5}{2 \pi V} 
\int d \hat {\bf k} \, \Lambda_{ij,lm} C^*_{ij} C_{lm} .
\ee

If the integration is performed over a large volume $V$ of the
Universe (in terms of $\beta^{-3}$) with $N$ bubbles, one obtains
in the quadrupole approximation
\be
\int d \hat {\bf k} \, \Lambda_{ij,lm} C^*_{ij} C_{lm} 
\propto N \beta^{-8} \propto \frac{V}{v_b^3 \beta^5},
\ee
and hence we recover the well known result~\cite{Kamionkowski:1993fg}
\be
\frac{E_{GW}}{E_{tot}} \propto
\kappa^2 v_b^3 \left( \frac{H}\beta \right) ^2 
\left( \frac{\alpha}{\alpha + 1} \right)^2.
\ee

The remainder of the present work is devoted to discussing numerical
results for the function $\Delta(x,v_b)$.

\section{Numerical Results}

To calculate the function $\Delta(x,v_b)$, we proceed as
follows. First, we choose the main parameters of the simulation: The
radius of the part of the Universe under consideration $L_U$, and the
number of directions to integrate over, $N_k$. Since the GW
production is isotropic, the directional variation indicates the
statistical uncertainty of the simulation.

Using this information, we simulate a scenario with random nucleation
times and locations for the bubbles following the nucleation
probability in eq.~(\ref{nuc_prob}).  As in
ref.~\cite{Kosowsky:1992vn}, our volume is spherical and the bubbles
are cut off by the boundary which is equivalent to having a mirror
symmetric configuration beyond the volume.

\begin{figure}[t]
\begin{center}
\includegraphics[width=0.95 \textwidth, clip]{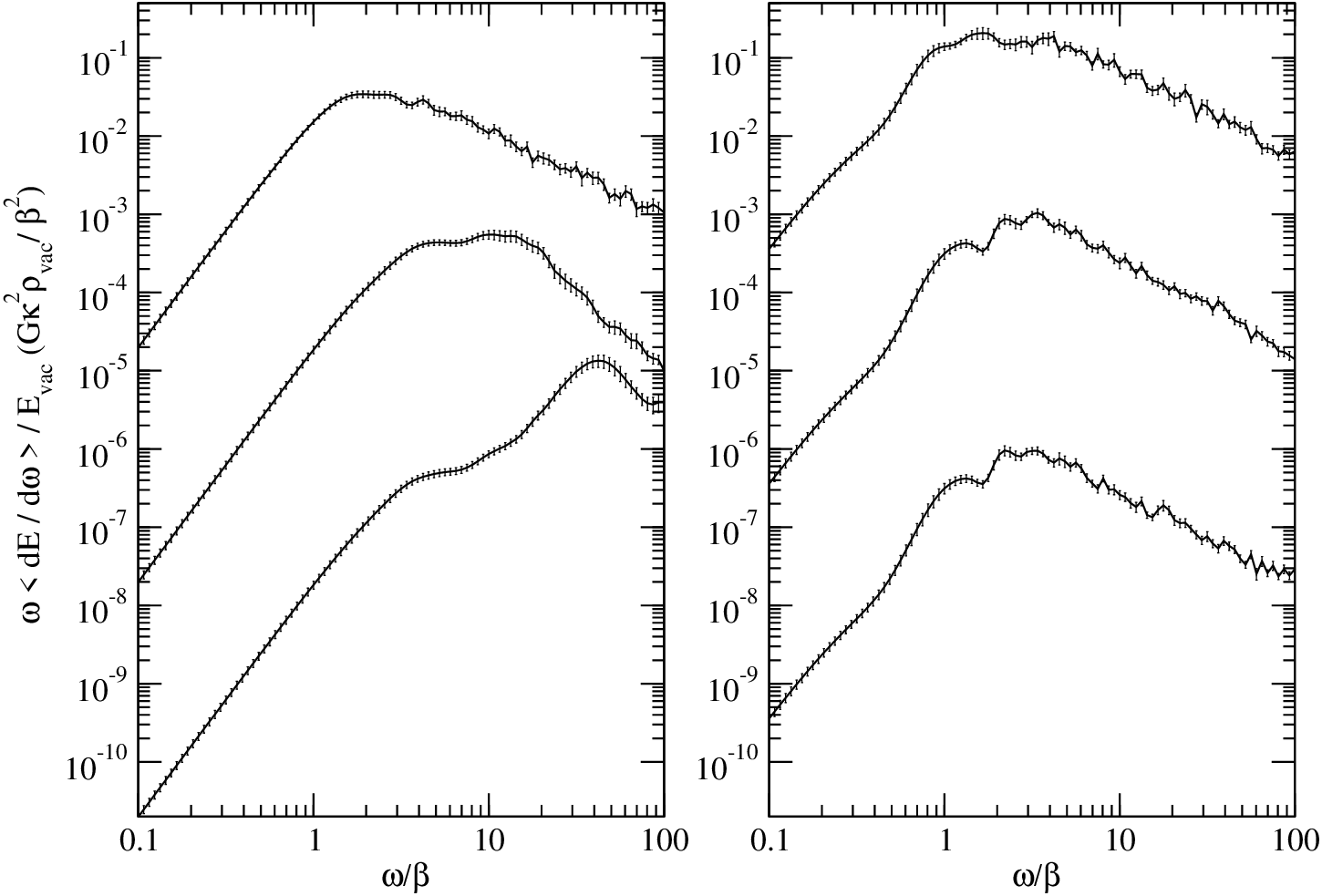}
\end{center}
\vskip -0.8cm
\caption{%
\label{fig_comp}
The left panel shows the fraction of gravitational radiation for a
small simulation, $L_U=3 v_b / \beta$, 7 bubbles. For small wall
velocities, significant boundary effects are visible. The right
simulation is relatively large, $L_U=7 v_b / \beta$, 109 bubbles. Both
simulations have been integrated over $N_k=32$ directions. Velocities
decrease from top to bottom as $v_b= \{1, \, 0.1, \, 0.01\}$.}
\end{figure}

The result for a small and a large Universe and a few directions,
$N_k=32$, is shown in Fig.~\ref{fig_comp}.  For large wall velocities,
the size of the simulated part of the Universe mostly influences the
total amplitude of the GW radiation. However, for smaller wall
velocities, the produced radiation at high frequencies depends
crucially on the size of the Universe. For small velocities the high
frequency part of the spectrum approaches the quadrupole
approximation, since in this case ${\bf k\cdot x}\to{\bf k\cdot
x_n}$. This is not identical to the quadrupole approximation (only the
phase in eq.~(\ref{eq_Aij}) vanishes), but it turns out that the
additional phase in eq.~(\ref{eq_Cij}) from the bubble position solely
leads to a slight suppression of the low frequency part of the
spectrum in our numerical results.

The enhancement of the high-frequency part in the quadrupole
approximation can be explained in the following way. After the
nucleation of the first bubble, the phase transition ends when the
whole space is in the new phase, which typically takes a time $\tau
\approx 5/\beta$. However, if the radius of the simulated volume $L_U$
is chosen too small, the simulation ends when the first nucleated
bubble covers the whole spherical volume. This leads to a suppression
of the total GW density and also to a rather steep edge in the time
profile that amplifies the high frequency part of the spectrum. This
enhancement effect is non-physical and just arises because of the
finite size of our test volume and the presence of the boundary. It
disappears for larger volumes, when the phase transition is finished
before the first nucleated bubble covers the whole spherical volume
and boundary effects are small. For large bubble velocities, when the
quadrupole approximation does not apply, the enhancement also
disappears, since effects from large bubbles are in general reduced
due to the fast oscillation of the integrand in
eq.~(\ref{eq_Aij}). This motivates our final choice $L_U = 7v_b/\beta$
in order to suppress these boundary effects.

The error bars in Fig.~\ref{fig_comp} result from the integration over
the $N_k=32$ different directions. If the number of directions would
be further increased, most of the displayed oscillatory features in
the spectrum would persist, since they result from the finiteness of
the size of the simulation $L_U$ and depend on the specific nucleation
positions and times of the randomly generated scenario. In addition,
the amplitude of the low frequency part of the spectrum and the total
emitted radiation depend sensitively on the times of the first bubble
nucleation and first collision, which are however quite similar in the
different scenarios. To decrease these features, one has to further
increase the size of the Universe and/or average over several
scenarios. We will do the latter in the following.

\begin{figure}[t]
\begin{center}
\includegraphics[width=0.95 \textwidth, clip]{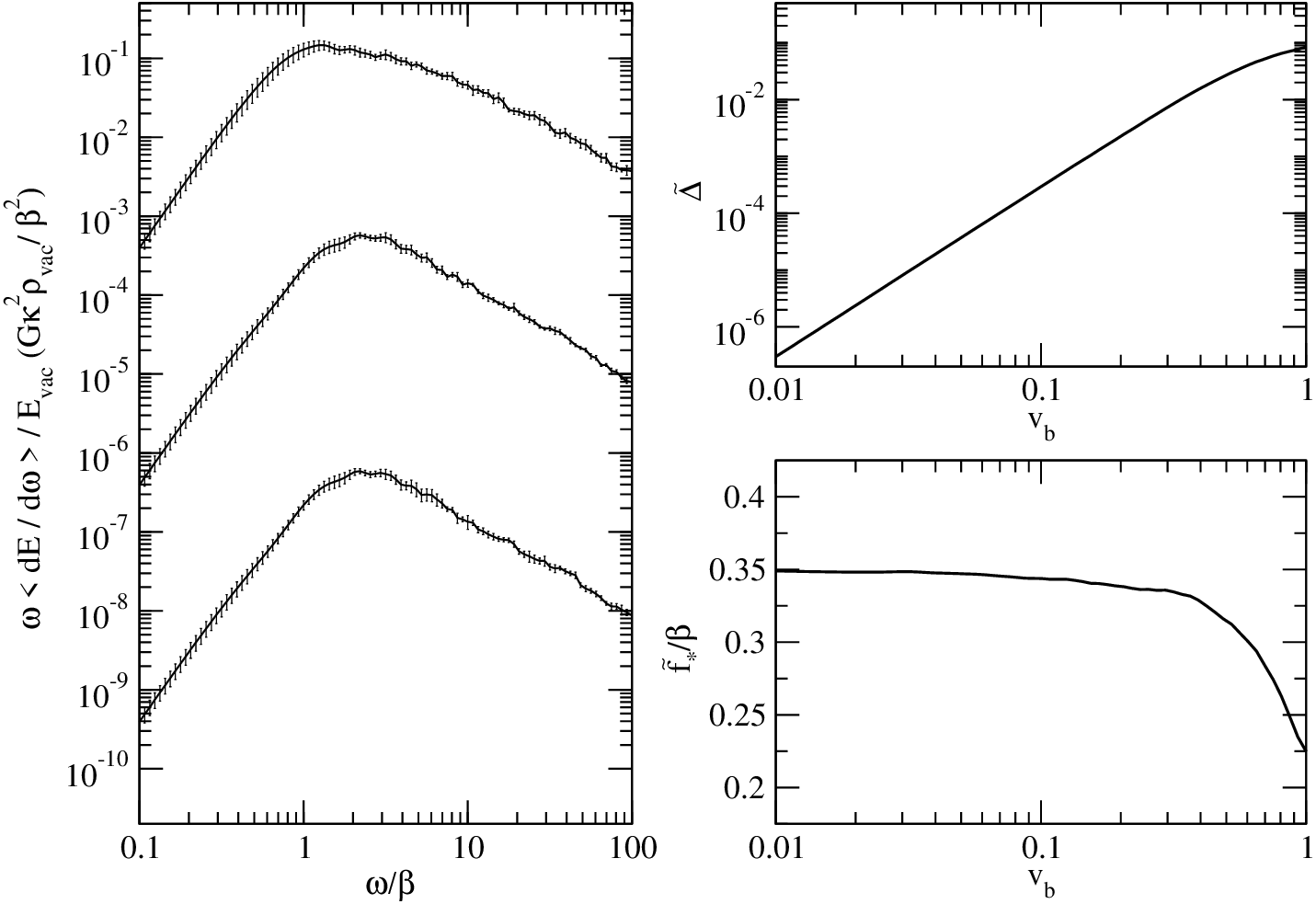}
\end{center}
\vskip -0.8cm
\caption{%
\label{fig_final}
The left panel shows the spectrum of gravitational radiation for a
simulation with $L_U=7 v_b / \beta$, integrated over $N_k=32$
directions. The error bars result from averaging over eight different
scenarios. Velocities decrease from top to bottom as $v_b= \{1, \,
0.1, \, 0.01\}$. The right panel shows the parameters $\tilde\Delta$
and $\tilde f_*/\beta$ as functions of $v_b$. }
\end{figure}

Our final results, shown in Fig.~\ref{fig_final}, are obtained for the
parameters
\be
L_U=7 v_b / \beta, \quad N_k=32, 
\ee
and averaged over eight scenarios. The given error bars result from
this averaging procedure. For very small frequencies, one expects
$\Omega_{GW*}$ to scale as $\omega^3$, since $C_{ij}(\omega)$ becomes
constant in this limit. This is reflected by the numerical results.
For large frequencies, we find that $\Omega_{GW*}$ scales as
$\omega^{-1}$, unlike the two-bubbles case that was discussed in
ref.~\cite{Kosowsky:1991ua}. We parametrize the spectrum as
\be
\label{res1}
\Omega_{GW*}(f_*) = \tilde \Omega_{GW*} 
\frac{(a+b) \, \tilde f_*^{b} f_*^a}{ b \tilde f_*^{(a+b)} + a f_*^{(a+b)}},
\ee
with the peak frequency $\tilde f_* = \tilde \omega / 2 \pi$ and the
peak amplitude $\tilde \Omega_{GW*}$ and both parameters are functions
of the wall velocity. The two exponents obtained by fitting lie in the
range $a \in [2.66, 2.82]$ and $b \in [0.90, 1.19]$. The most
interesting case from a phenomenological point of view is given by
large wall velocities, $v_b \approx 1$, in which case the fit yields
$a \approx 2.8$ and $b \approx 1.0$. Note that the fit is optimized
for a frequency range close to the peak frequency, and does not
correctly reproduce the asymptotic low frequency behavior.

The main result of our analysis is that the spectrum scales for high
frequencies close to $\omega^{-1}$ in contrast to the two bubble case
presented in ref.~\cite{Kosowsky:1991ua}. This feature is already
present for a rather small volume (just a few bubbles) and high wall
velocities ($v_b \simeq 1$) as depicted in the left panel of
Fig.~\ref{fig_comp}. The reason for this qualitative difference
between the results is hence probably the time-dependent cutoff
function that has been used in ref.~\cite{Kosowsky:1991ua} in order to
terminate the phase transition.

 The spectrum that would be observed today is
obtained by red-shifting this result according to
\bea
\label{res2}
\tilde f &=& 16.5 \times 10^{-3} \textrm{mHz} \left( \frac{\tilde f_*}{\beta} \right) 
\left( \frac{\beta}{H_*} \right)
\left( \frac{T_*}{100 \textrm{ GeV}}\right) \left( \frac{g_*}{100}\right)^{1/6}, \\
h^2 \tilde \Omega_{GW} &=& 1.67 \times 10^{-5} \,\, \tilde \Omega_{GW*}
\left( \frac{100}{g_*}\right)^{1/3}  \nn \\
&=& 1.67 \times 10^{-5} \tilde \Delta \,
\kappa^2 \left( \frac{H}{\beta} \right)^2 
\left( \frac{\alpha}{\alpha + 1} \right)^2
\left( \frac{100}{g_*}\right)^{1/3}.
\eea
The two functions $\tilde f_*/\beta$ and $\tilde \Delta$ are displayed
in the right panels of Fig.~\ref{fig_final} and are approximately
given by
\bea
\tilde \Delta &=& \frac{0.11 \, v_b^3}{0.42 + v_b^2}, \\
\tilde f_*/\beta &=& \frac{0.62}{1.8 -0.1 v_b + v_b^2}.
\label{res3}
\eea
Notice that the peak frequency and amplitude agree reasonably well
with the results presented in ref.~\cite{Kamionkowski:1993fg} (our
peak amplitude is about $50\%$ larger for $v_b \approx 1$, while for
small $v_b$ both results agree within the statistical errors). In the
light of the analysis presented in ref.~\cite{Caprini:2006rd}, the
dependence of the peak frequency on the wall velocity has the
following physical interpretation. For small velocities, $v_b \ll 1$,
the phase transition lasts long compared to the relevant distance
scale that is given by the average bubble size. In this case, the GW
spectrum inherits the time scale of the source, $\tilde f \sim
\beta$. If one used in eqs.~(\ref{eq_Tij})-(\ref{eq_Aij}) a
wall velocity much larger than the speed of light, $v_b \gg 1$, the
phase transition would be very short compared to the relevant distance
scale and the GWs would inherit the distance scale of the source,
$\tilde f \sim
\beta / v_b$. This effect leads to a decrease in the peak frequency in
the transition region where the wall velocity is close to the speed of
light.

\begin{figure}[t]
\begin{center}
\includegraphics[width=0.95 \textwidth, clip]{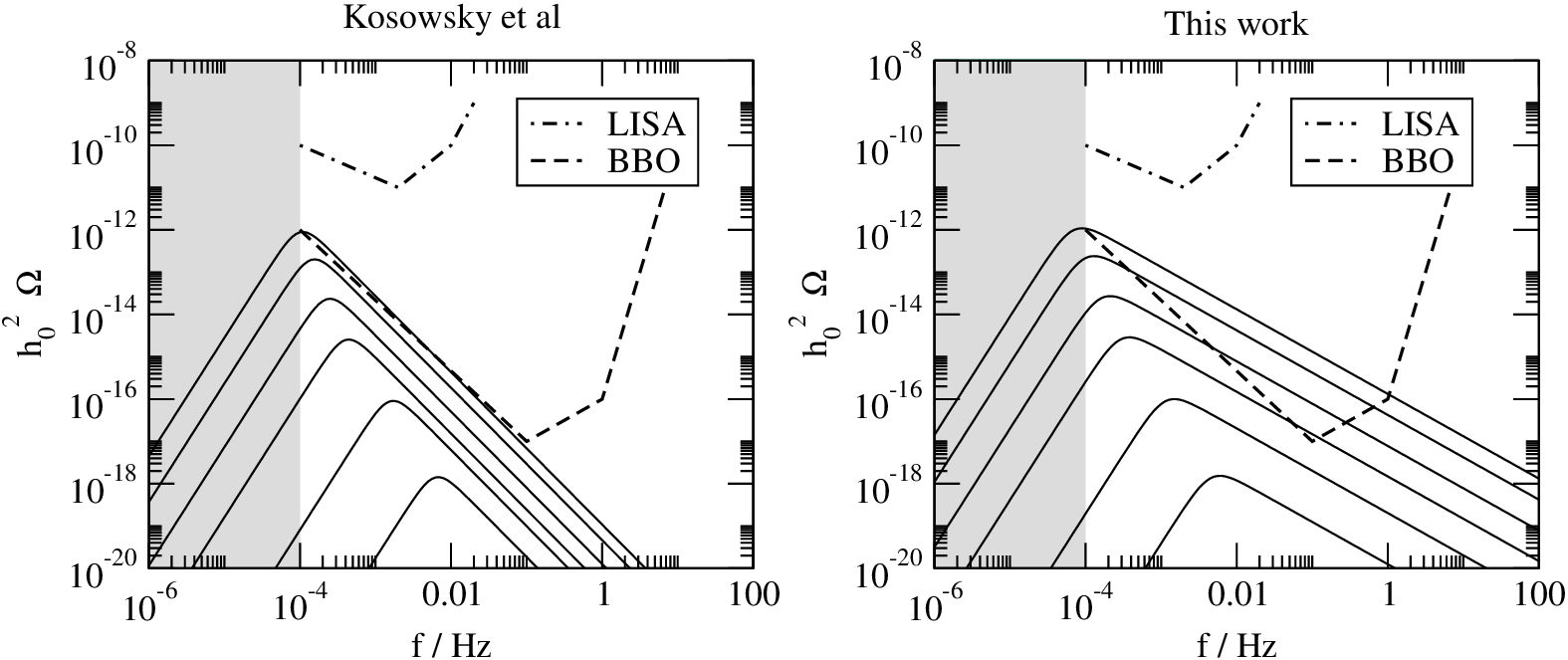}
\end{center}
\vskip -0.8cm
\caption{%
Several spectra of gravitational radiation according to the old and
new formulas. The parameters are taken from ref.~\cite{KonHub} and
given in table \ref{tab_spectrum} with $\alpha$ decreasing from top to
bottom. In the shaded region, the sensitivity of LISA and BBO is
expected to drop considerably.}
\label{fig_spectrum}
\end{figure}

\begin{table}
\begin{tabular}[b]{|c||c|c|c|}
\hline
\quad set \quad&
\quad $\alpha$ \quad\quad &
\,\, $\beta/H$ \quad &
\,\, $T_*$ / GeV \quad  \\
\hline
\hline
1 & 0.03 & 1000 & 130 \\
\hline
2 & 0.05 & 300 & 110 \\
\hline
3 & 0.07 & 100 & 85 \\
\hline
4 & 0.1  & 60 & 80 \\
\hline
5 & 0.15 & 40 & 75 \\
\hline
6 & 0.2  & 30 & 70 \\
\hline
\end{tabular}
\caption{Sets of parameters used in Fig. \ref{fig_spectrum}.
\label{tab_spectrum}
}
\end{table}

\section{Conclusions}

We reexamined the spectrum of gravitational wave radiation generated
by bubble collisions during a first-order phase transition in the
envelope approximation.  Using refined numerical simulations, our main
finding is that the spectrum falls off only as $f^{-1.0}$ at high
frequencies, considerably slower than appreciated in the literature.
This behavior is most probably related to the many small bubbles
nucleated at a later stage of the phase transition~\footnote{The
effect has been mentioned in refs.~\cite{Kosowsky:1992rz,
Kosowsky:1992vn, Kamionkowski:1993fg}, but could never be clarified
because of numerical uncertainties.}. This result is especially
interesting in the light of recent
investigations~\cite{Grojean:2006bp, KonHub} that indicate that in the
case of a first-order electroweak phase transition (obtained by a
singlet sector~\cite{Huber:2000mg, Espinosa:2007qk} or higher
dimensional operators~\cite{Grojean:2004xa, Bodeker:2004ws,
Delaunay:2007wb}) the peak frequency of the produced radiation is
below the best sensitivity range of planed satellite experiments, such
as LISA and BBO~\cite{Danzmann:2003tv, Corbin:2005ny}. This effect is
shown in Fig.~\ref{fig_spectrum} for several typical parameter sets
for the phase transition in the nMSSM~\cite{KonHub}. Notice that the
discussion in ref.~\cite{KonHub} suggests that stronger phase
transitions in general lead to smaller peak frequencies due to a
decrease in the parameters $\beta/H$ and $T_*$. This amplifies the
importance of the high frequency part of the gravitational wave
spectrum. Notice also that a flatter spectrum simplifies the
distinction from other sources of stochastic gravitational waves, such
as turbulence~\cite{Kosowsky:2001xp, Dolgov:2002ra, Caprini:2006jb,
Gogoberidze:2007an} or preheating after
inflation~\cite{Khlebnikov:1997di, Easther:2006gt, Easther:2006vd,
GarciaBellido:2007dg, GarciaBellido:2007af, Dufaux:2007pt}. Besides,
we found that the peak frequency slightly depends on the expansion
velocity of the bubbles and decreases for higher wall velocities. Our
quantitative results are summarized by eqs.~(\ref{res1})-(\ref{res3}).

Finally, we would like to comment on the recent paper
\cite{Caprini:2007xq}, where an analytic approach to the GW
production by collisions based on stochastic considerations was
presented. In this approach, assumptions have to be made about the
time-dependence of unequal time correlations of the velocity field. In
their favored model, the authors obtain a scaling as $\omega^{-2}$ for
the high frequency part of the spectrum.  We suspect that this
disparity is due to conceptual differences.

First, notice that the treatment presented here is based on two main
ingredients: The thin wall and the envelope approximations. Even
though the stochastic approach in ref.~\cite{Caprini:2007xq} does not
require the thin wall approximation, the results are also valid in
this limit, such that this approximation cannot be responsible for the
different spectra. However, in the approach of
ref.~\cite{Caprini:2007xq} the collided and uncollided regions of the
bubbles are treated equally but in a stochastic manner. Breaking of
the spherical symmetry, necessary for GW production, is encoded in
assumptions on the velocity correlation functions.  This is in
contrast to our analysis in which the well tested
\cite{Kosowsky:1992rz} envelope approximation breaks the spherical
symmetry in a realistic way.

Furthermore, only the time dependent mean bubble radius enters into
the stochastic calculation, while in the analysis at hand bubbles with
a realistic size distribution are simulated. The occurrence of many
small bubbles probably enhances the high frequency part of the spectrum, as
already argued in refs.~\cite{Kosowsky:1992rz, Kosowsky:1992vn,
Kamionkowski:1993fg}.

Finally, it is interesting to see that even in the stochastic
treatment the high frequency part of the spectrum can scale as
$\omega^{-1}$ (in agreement with our results) if one assumes that the
source is fully uncorrelated at unequal times. Especially at late
stages of the phase transition, many small bubbles are generated,
start to collide and are absorbed by neighboring bubbles at a large
rate such that in this regime the assumption of non-correlation might
indeed be plausible.

\section*{Acknowledgments}
We would like to thank G.~Servant for helpful discussions on
ref.~\cite{Caprini:2007xq}.  T.K. is supported by the Swedish Research
Council (Vetenskapsr{\aa}det), Contract No.~621-2001-1611 and by the
EU FP6 Marie Curie Research \& Training Network 'UniverseNet'
(MRTN-CT-2006-035863).

\appendix
\section{Some details about the numerical algorithm}

In this section we will present an efficient way of evaluating the
integrals given in eqs.~(\ref{eq_Tij})-(\ref{eq_Aij}). It turns out
that rotating the vector $\hat {\bf k}$ parallel to the z-axis is very
useful.  One advantage is that the projection of the
transverse-traceless part simplifies to
\be
\Lambda_{ij,lm} T^*_{ij} T_{lm} = \frac12 (T^*_{xx} - T^*_{yy}) (T_{xx} - T_{yy})
+  T^*_{xy} T_{xy} + T^*_{yx} T_{yx}. 
\ee
Using cylindrical coordinates and denoting the rotated uncollided region as 
$S^\prime_n$, this leads to
\bea
\label{def_dE}
\frac{dE_{GW}}{d\omega d\Omega} &=& 4 G \rho^2_{vac} \kappa^2 
v_b^6 \omega^2  ( |C_+|^2 + |C_-|^2 ), \\
\label{def_C}
C_\pm (\omega) &=& \frac{1}{6 \pi}
\sum_n \int dt \, e^{ i \omega ( t - z_n )} 
 (t-t_n)^3 \, A_{n,\pm}(\omega, t), \\
\label{def_Apm}
A_{n,\pm} (\omega,t) &=& \int_{-1}^1 dz 
\, e^{- i v_b \omega (t-t_n) z } B_{n,\pm}(z,t), 
\eea
\be
\label{def_Bpm}
B_{n,+}(z,t) = \frac{(1 - z^2)}2 \int_{S^\prime_n} d\phi \cos(2\phi), \quad 
B_{n,-}(z,t) = \frac{(1 - z^2)}2 \int_{S^\prime_n} d\phi \sin(2\phi). 
\ee
The second advantage of this coordinate system is that the integrals
$B_{n,\pm}(z)$ do not depend on the frequency $\omega$ or the wall
velocity $v_b$. Hence, the functions $B_{n,\pm}(z)$ need only to be
determined once for all values of $\omega$ and $v_b$ what greatly
improves the performance of the numerical evaluation of the integrals.

Notice that the function $A_{n,ij}(\omega, t)$ in eq.~(\ref{def_Apm})
does not depend on the frequency $\omega$ and the wall velocity $v_b$
separately, but only on the product $\omega v_b$. Hence, we choose an
exponential distribution for the values of the wall velocity $v_b$ and
the frequency $\omega$ to reduce the number of necessary evaluations
of $A_{n,\pm}(\omega, t)$. For each direction, we then evaluate the
integrals (\ref{def_dE})-(\ref{def_Bpm}) and integrate over the
different directions. 

\begin{figure}[t]
\begin{center}
\includegraphics[width=0.8 \textwidth, clip]{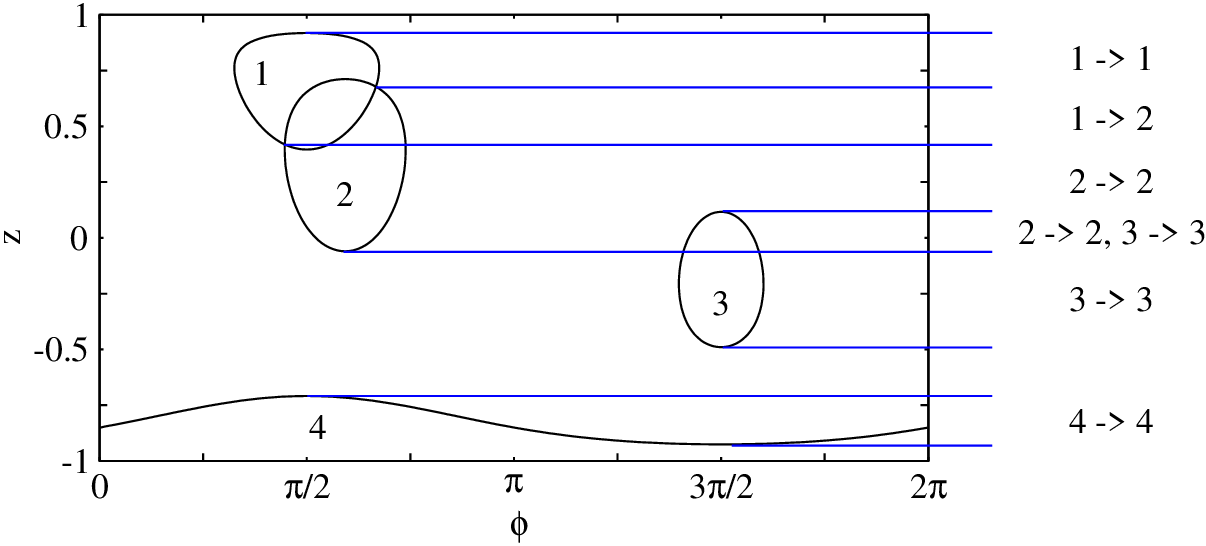}
\end{center}
\vskip -0.8cm
\caption{%
\label{fig_alg}
The figure depicts the case of a bubble that intersects with four
neighboring bubbles. The data structure we use contains the
information which bubbles constitute the boundaries of the uncollided
regions in each segment. This improves the performance greatly
especially at the end of the phase transition when virtually all
bubbles intersect with each other but only a few bubbles are relevant
in each segment.  }
\end{figure}

Finally, we would like to comment on the numerical accuracy of our
simulation. The integration in eq.~(\ref{def_Bpm}) is performed by
determining the intersections of the overlapping bubbles, while the
integration in eq.~(\ref{def_Apm}) is done on an adaptive grid such
that the relative error never exceeds $10^{-5}$. This requires for the
highest frequencies up to $10^4$ evaluations of the functions
$B_{n,\pm}(z)$. In order to improve the performance we employ a data
structure that for a fixed time $t$ and a specific bubble $n$ encodes
the regions that are uncollided and do not intersect with other
bubbles. The $z$-space is divided into different segments, where in
every segment the data structure contains the information which
bubbles constitutes the borders of the intersecting regions. A
graphical representation of this data structure is depicted in
Fig.~\ref{fig_alg}. This is especially useful at late stages of the
phase transition, when almost all bubbles intersect with each other,
but only a few bubbles are relevant for each segment. The dominant
error in the final spectrum results from the fact that the time
integration in eq.~(\ref{def_C}) is done on an equidistant grid with
size $N_t=512$. We checked that the error resulting from this choice
never exceeds a few percent for the presented results. As a first
test, we used this algorithm to reproduce the results in the two
bubble case and with a cutoff function as presented in
ref.~\cite{Kosowsky:1992vn}.

A final source of uncertainty is the finite size of the volume. As
argued in the main text, the completion of the phase transition takes
approximately the time $\tau \approx 5 / \beta$ and the the size of
the volume under consideration should be chosen accordingly. It turns
out that for the portion of the spectrum under consideration, a size
of $L_U = 7 v_b / \beta$ is sufficient to suppress boundary effects,
as long as the first bubble nucleates not too far from the center of
the volume which we assure in our simulations.


\begin{thebibliography}{99}
\bibliographystyle{unsrt}

\bibitem{Witten:1984rs}
  E.~Witten,
  ``Cosmic Separation Of Phases,''
  Phys.\ Rev.\  D {\bf 30} (1984) 272.

\bibitem{Hogan}
  C.~Hogan,
  ``Gravitational Radiation from Cosmological Phase Transitions,''
  Mon.\ Not.\ R.\ Astron.\ Soc.\ {\bf 218} (1986) 629.

\bibitem{Kosowsky:1991ua}
  A.~Kosowsky, M.~S.~Turner and R.~Watkins, ``Gravitational Radiation
  From Colliding Vacuum Bubbles,'' Phys.\ Rev.\ D {\bf 45} (1992)
  4514.  

\bibitem{Kosowsky:1992rz}
  A.~Kosowsky, M.~S.~Turner and R.~Watkins,
  ``Gravitational waves from first order cosmological phase transitions,''
  Phys.\ Rev.\ Lett.\  {\bf 69} (1992) 2026.

\bibitem{Kosowsky:1992vn}
  A.~Kosowsky and M.~S.~Turner,
  ``Gravitational Radiation From Colliding Vacuum Bubbles: Envelope
  Approximation To Many Bubble Collisions,''
  Phys.\ Rev.\  D {\bf 47} (1993) 4372
  [arXiv:astro-ph/9211004].

\bibitem{Kamionkowski:1993fg}
  M.~Kamionkowski, A.~Kosowsky and M.~S.~Turner,
  ``Gravitational radiation from first order phase transitions,''
  Phys.\ Rev.\  D {\bf 49} (1994) 2837
  [arXiv:astro-ph/9310044].


\bibitem{Grojean:2006bp}
  C.~Grojean and G.~Servant,
  ``Gravitational waves from phase transitions at the electroweak scale and
  beyond,''
  Phys.\ Rev.\  D {\bf 75} (2007) 043507
  [arXiv:hep-ph/0607107].

\bibitem{KonHub}
  S.~J.~Huber and T.~Konstandin,
  ``Production of Gravitational Waves in the nMSSM,''
  JCAP {\bf 0805} (2008) 017
  [arXiv:0709.2091 [hep-ph]].

\bibitem{Kahniashvili:2008pf}
  T.~Kahniashvili, A.~Kosowsky, G.~Gogoberidze and Y.~Maravin,
  ``Detectability of Gravitational Waves from Phase Transitions,''
  Phys.\ Rev.\  D {\bf 78} (2008) 043003
  [arXiv:0806.0293 [astro-ph]].

\bibitem{Weinbergbook}
  S.~Weinberg,
  ``Gravitation and Cosmology,''
  Wiley, New York (1982).

\bibitem{Steinhardt:1981ct}
  P.~J.~Steinhardt,
  ``Relativistic Detonation Waves And Bubble Growth In False Vacuum Decay,''
  Phys.\ Rev.\ D {\bf 25}, 2074 (1982).

\bibitem{Laine:1993ey}
  M.~Laine,
  ``Bubble Growth As A Detonation,''
  Phys.\ Rev.\  D {\bf 49} (1994) 3847
  [arXiv:hep-ph/9309242].

\bibitem{Caprini:2006rd}
  C.~Caprini, R.~Durrer and R.~Sturani,
  ``On the frequency of gravitational waves,''
  Phys.\ Rev.\  D {\bf 74}, 127501 (2006)
  [arXiv:astro-ph/0607651].

\bibitem{Huber:2000mg}
  S.~J.~Huber and M.~G.~Schmidt,
  ``Electroweak baryogenesis: Concrete in a SUSY model with a gauge  singlet,''
  Nucl.\ Phys.\  B {\bf 606} (2001) 183
  [arXiv:hep-ph/0003122].

\bibitem{Espinosa:2007qk}
  J.~R.~Espinosa and M.~Quiros,
  ``Novel effects in electroweak breaking from a hidden sector,''
  Phys.\ Rev.\  D {\bf 76} (2007) 076004
  [arXiv:hep-ph/0701145].

\bibitem{Grojean:2004xa}
  C.~Grojean, G.~Servant and J.~D.~Wells,
  ``First-order electroweak phase transition in the standard model with a  low
  cutoff,''
  Phys.\ Rev.\  D {\bf 71} (2005) 036001
  [arXiv:hep-ph/0407019].

\bibitem{Bodeker:2004ws}
  D.~Bodeker, L.~Fromme, S.~J.~Huber and M.~Seniuch,
  ``The baryon asymmetry in the standard model with a low cut-off,''
  JHEP {\bf 0502}, 026 (2005)
  [arXiv:hep-ph/0412366].

\bibitem{Delaunay:2007wb}
  C.~Delaunay, C.~Grojean and J.~D.~Wells,
  ``Dynamics of Non-renormalizable Electroweak Symmetry Breaking,''
  JHEP {\bf 0804} (2008) 029
  [arXiv:0711.2511 [hep-ph]].

\bibitem{Danzmann:2003tv}
  K.~Danzmann and A.~Rudiger,
  ``Lisa Technology - Concept, Status, Prospects,''
  Class.\ Quant.\ Grav.\  {\bf 20} (2003) S1.

\bibitem{Corbin:2005ny}
  V.~Corbin and N.~J.~Cornish,
  ``Detecting the cosmic gravitational wave background with the big bang
  observer,''
  Class.\ Quant.\ Grav.\  {\bf 23} (2006) 2435
  [arXiv:gr-qc/0512039].

\bibitem{Kosowsky:2001xp}
  A.~Kosowsky, A.~Mack and T.~Kahniashvili,
  ``Gravitational radiation from cosmological turbulence,''
  Phys.\ Rev.\  D {\bf 66}, 024030 (2002)
  [arXiv:astro-ph/0111483].

\bibitem{Dolgov:2002ra}
  A.~D.~Dolgov, D.~Grasso and A.~Nicolis,
  ``Relic backgrounds of gravitational waves from cosmic turbulence,''
  Phys.\ Rev.\ D {\bf 66}, 103505 (2002)
  [arXiv:astro-ph/0206461].

\bibitem{Caprini:2006jb}
  C.~Caprini and R.~Durrer,
  ``Gravitational waves from stochastic relativistic sources: Primordial
  turbulence and magnetic fields,''
  Phys.\ Rev.\  D {\bf 74} (2006) 063521
  [arXiv:astro-ph/0603476].

\bibitem{Gogoberidze:2007an}
  G.~Gogoberidze, T.~Kahniashvili and A.~Kosowsky,
  ``The spectrum of gravitational radiation from primordial turbulence,''
  Phys.\ Rev.\  D {\bf 76} (2007) 083002
  [arXiv:0705.1733 [astro-ph]].

\bibitem{Khlebnikov:1997di}
  S.~Y.~Khlebnikov and I.~I.~Tkachev,
  ``Relic gravitational waves produced after preheating,''
  Phys.\ Rev.\  D {\bf 56}, 653 (1997)
  [arXiv:hep-ph/9701423].

\bibitem{Easther:2006gt}
  R.~Easther and E.~A.~Lim,
  ``Stochastic gravitational wave production after inflation,''
  JCAP {\bf 0604}, 010 (2006)
  [arXiv:astro-ph/0601617].

\bibitem{Easther:2006vd}
  R.~Easther, J.~T.~Giblin and E.~A.~Lim,
  ``Gravitational Wave Production At The End Of Inflation,''
  Phys.\ Rev.\ Lett.\  {\bf 99}, 221301 (2007)
  [arXiv:astro-ph/0612294].

\bibitem{GarciaBellido:2007dg}
  J.~Garcia-Bellido and D.~G.~Figueroa,
  ``A stochastic background of gravitational waves from hybrid preheating,''
  Phys.\ Rev.\ Lett.\  {\bf 98} (2007) 061302
  [arXiv:astro-ph/0701014].

\bibitem{GarciaBellido:2007af}
  J.~Garcia-Bellido, D.~G.~Figueroa and A.~Sastre,
  ``A Gravitational Wave Background from Reheating after Hybrid Inflation,''
  Phys.\ Rev.\  D {\bf 77}, 043517 (2008)
  [arXiv:0707.0839 [hep-ph]].

\bibitem{Dufaux:2007pt}
  J.~F.~Dufaux, A.~Bergman, G.~N.~Felder, L.~Kofman and J.~P.~Uzan,
  ``Theory and Numerics of Gravitational Waves from Preheating after
  Inflation,''
  Phys.\ Rev.\  D {\bf 76}, 123517 (2007)
  [arXiv:0707.0875 [astro-ph]].

\bibitem{Caprini:2007xq}
  C.~Caprini, R.~Durrer and G.~Servant,
  ``Gravitational wave generation from bubble collisions in first-order phase
  transitions: an analytic approach,''
  Phys.\ Rev.\  D {\bf 77}, 124015 (2008)
  [arXiv:0711.2593 [astro-ph]].



\end{thebibliography}
\end{document}